\documentclass[final]{aipproc}
\usepackage{sidecap,graphicx,boxedminipage,epsfig,wrapfig,floatflt,shadow}

\layoutstyle{8x11double}

\begin{document}

\title{Using Reflection with Peers to Help Students Learn Effective Problem Solving Strategies}
\author{Andrew Mason and Chandralekha Singh}{
  address={Department of Physics and Astronomy, University of Pittsburgh, Pittsburgh, PA, 15260}
}


\begin{abstract}
We describe a study in which introductory physics students engage in reflection with peers about problem solving.
The recitations for an introductory physics course 
with 200 students 
were broken into the ``Peer Reflection" (PR) group
and the traditional group.
Each week in recitation, students in the PR group reflected in small teams on selected problems from the homework.
The graduate and undergraduate teaching assistants (TAs) in the 
PR group recitations provided guidance and coaching to help students learn effective problem solving heuristics. 
In the recitations for the traditional group, 
students had the opportunity to ask the graduate TA questions about the homework before they took a weekly quiz. 
On the final exam with only multiple-choice questions, the PR group drew diagrams on 
more problems than the traditional group, even when there was no external reward for doing so. 
Since there was no partial credit for drawing the diagrams on the scratch books, students did not draw diagrams simply to get 
credit for the effort shown and must value the use of diagrams for solving problems if they drew them. 
We also find that, regardless of whether the students belonged to the traditional or PR groups, those who drew more diagrams
for the multiple-choice questions outperformed those who did not draw them.\\


\end{abstract}

\maketitle

\section{Introduction}

\vspace{-.03in}

Reflection is an integral component of effective problem solving~\cite{edit1}. While experts in a particular field
reflect and exploit problem solving as an opportunity for organizing and extending their knowledge, students often
need feedback and support to learn how to use problem solving as an opportunity for learning. 
Our prior research has shown that, even with minimal guidance from the instructors, students can benefit from peer interaction~\cite{singhgroup}.
Those who worked with peers not only outperformed an equivalent group of students who worked alone on the same task, 
but collaboration with a peer led to ``co-construction" of knowledge in $29\%$ of the cases~\cite{singhgroup}. 

Here, we describe a study in which algebra-based introductory physics students in the Peer Reflection group (PR group) 
were provided guidance and support to reflect upon problem solving with peers and undergraduate and graduate teaching assistants 
in the recitation class~\cite{mason}. 

\vspace{-0.15in}
\section{Methodology}
\vspace{-0.09in}

The investigation involved an introductory algebra-based physics course mostly taken by students 
with interest in health related professions.  The course had 200 students and was broken into two sections both of which
met on Tuesdays and Thursdays and were taught by the same professor who had taught both sections of the course before. 
A class poll at the beginning of the course indicated that more than $80\%$ of the students had taken at least one physics course in 
high school, and perhaps more surprisingly, more than $90\%$ of the students had taken at least one calculus course (although the 
physics course in which they were enrolled was an algebra-based course).

The daytime section taught during the day was the traditional group and had 107 students whereas the evening section called the
``Peer Reflection" group or PR group had 93 students. 
The lectures, all homework assignments, the midterm exams and the final exam were identical for the daytime and evening sections of the course.
Moreover, the instructor emphasized effective problem solving strategies, e.g., performing a conceptual
analysis of the problem and planning of the solution before implementing the plan and importance of evaluating the solution throughout the semester
in both the traditional and peer-reflection groups. 

Each week, students in both groups were supposed to turn in answers to the assigned homework problems (based upon the material covered in
the previous week) using an online homework system for some course credit. In addition, students in both groups were supposed to submit
a paper copy of the homework problems which had the details of the problem solving approach at the
end of the recitation class to the TA for some course credit. While the online homework solution was graded for correctness, the TA only 
graded the paper copies of the submitted homework for completeness on a three point scale (full score, half score or zero).

The weighting of each component of the course, e.g., midterm exams, final exam, class participation, homework and the scores allocated for the recitation 
were the same for both classes. Also, as noted earlier, all components of the course were identical for both groups except
the recitations which were conducted very differently for the PR and traditional groups.  
Although the total course weighting assigned to the recitations was the same for both groups (since all the other components of the course had the 
same weighting for both groups), the scoring of the recitations was different for the two groups. 
Students were given credit for attending recitation in both groups. 
Attendance was taken in the recitations using clickers for both the traditional group and the PR group. 
The traditional group recitations were traditional in which the TA would solve selected assigned homework problems on the blackboard
and field questions from students about their homework before assigning a quiz in the last 20 minutes of the recitation class. 
The recitation quiz problems given to the traditional group were similar to the homework problems selected for ``peer 
reflection" in the PR group recitations (but the quiz problems were not identical to the homework problems to discourage students in the 
traditional group from memorizing the answers to homework in preparation for the quiz). 
Students in the PR group reflected on three homework problems in each recitation class but no recitation quiz was given to the 
students in this group at the end of the recitation classes, unlike the traditional group, primarily due to the time constraints.
The recitation scores for the PR group students were assigned based mostly on the recitation attendance except students obtained bonus points for
helping select the ``best" student solution as described below.
Since the recitation scoring was done differently for the traditional and PR groups, the two groups
were curved separately so that the top $50\%$ in each group obtained A and B grades in view of the departmental policy.

As noted earlier, both recitation sections for the evening section (93 students total) together formed the 
PR group. The PR group intervention was based upon a field-tested cognitive 
apprenticeship model~\cite{cog1} of learning involving modeling, coaching, and fading to help students learn effective 
problem solving heuristics. In this approach, ``modeling" means that the TA demonstrates and exemplifies the effective problem solving 
skills that the students should learn. ``Coaching" means providing students opportunity to practice problem solving skills
with appropriate guidance so that they learn the desired skills.  
``Fading" means decreasing the feedback gradually with a focus on helping students develop self-reliance.

The specific strategy used by the students in the PR group involved
reflection upon problem solving with their peers in the recitations, while the TA and the undergraduate teaching assistants (UTAs)
exemplified the effective problem solving heuristics. The UTAs were chosen from those undergraduate students who had earned an $A^+$ grade in
an equivalent introductory physics course previously. The UTAs had to attend all the lectures in the semester in which they were UTAs for
a course and they communicated with the TA each week (and periodically with the course instructor) to determine the plan for the recitations. 
We note that, for effective implementation of the PR method, two UTAs were present in each recitation class along with the TA. These 
UTAs helped the TA in demonstrating and helping students to learn effective problem solving heuristics. 

In our intervention, each of the three recitation sections in the traditional group had about 35-37 students. 
The two recitations for the PR group had more than 40 students each (since the PR group was the evening section of the course, it was logistically not possible to break this group into three recitations). At the 
beginning of each PR recitation, students were asked to form nine teams of three to six
students chosen at random by the TA (these teams were generated by a computer program each week). 
The TA projected the names of the team members on the screen so that they could sit together at the beginning 
of each recitation class. Three homework questions were chosen for a particular recitation.  
The recitations for the two sections were coordinated by the TAs so that the recitation quiz problems given to the traditional group were based upon 
the homework problems selected for ``peer reflection" in the PR group recitations. Each of the three ``competitions" was carefully timed to take 
approximately 15 minutes, in order for the entire exercise to fit into the allotted fifty-minute time slot.

After each question was announced to the class, each of the nine teams were given three minutes to identify the ``best" solution by comparing and 
discussing among the group members.  If a group had difficulty coming up with a ``winner", the TA/UTA would intervene and facilitate the process.  
The winning students were asked to come to the front of the room, where they were assembled into three second-round groups.  
The process was repeated, producing three finalists. These students handed in their homework solutions to the TAs, after which the 
TA/UTA evaluation process began. 

The three finalists' solutions were projected one at a time on a screen using a web cam and computer projector. Each of the three panelists (the TA
and two UTAs) gave 
their critique of the solutions, citing what each of the finalists had done well and what could be done to further enhance the problem solving 
methodology in each case.  In essence, the TA and UTAs were ``judges" similar to the judges in the television show ``American Idol" and gave their 
``critique" of each finalist's problem solving performance. After each solution had been critiqued by each of the panelists, the students, using 
the clickers, voted on the ``best" solution. The TA and UTAs did not participate in the voting process. 

In order to encourage each team in the PR group to select the student with the most effective problem solving strategy as the winner for each 
problem, all students from the teams whose member advanced to the final round to ``win" the ``competition" were given course credit (bonus points). 
In particular, each of these team members (consolation prize winners) earned one third of the course credit given to the student whose solution was declared to be the ``winner".
This reward system made the discussions lively and the teams 
made good effort to advance the most effective solution to the next stage. 

While we video-taped a portion of the recitation class discussions when students reflected with peers, a good account of the effectiveness and 
intensity of the team discussions came from the TA and UTAs who generally walked around from team to team listening to the discussions but not 
interrupting the team members involved in the discussions unless facilitation was necessary for breaking a gridlock. 
The course credit and the opportunity to have the finalists' solutions voted on by the whole class encouraged students to argue passionately about 
the aspects of their solutions that displayed effective problem solving strategies. Students were constantly arguing about why drawing a diagram, 
explicitly thinking about the knowns and target variables, and explicitly justifying the physics principles that would be useful before writing the
equations are effective problem solving strategies.

Furthermore, the ``American Idol" style recitation allowed the TAs to discuss and convey to students in much more detail what solution styles were 
preferred and why.  Students were often shown what kinds of solutions were easier to read and understand, and which were more amenable to 
error-checking.  Great emphasis was placed on consistent use of notation, setting up problems through the use of symbols to define physical 
quantities, and the importance of diagrams in constructing solutions.

At the end of the semester, all of the students were given a final exam consisting of 40 multiple choice questions, 20 of which were primarily 
conceptual in nature and 20 of which were primarily quantitative (students had to solve a numerical or symbolic problem for a target quantity). 
Although the final exam was all multiple-choice, a novel assessment method was used. While students knew that the only thing that counted
for their grade was whether they chose the correct option for each multiple-choice question, each student was given an exam notebook which
he/she could use for scratchworks. 
We hypothesized that even if the final exam questions were in the multiple-choice format, students who value effective problem
solving strategies will take the time to draw more diagrams and do more scratchworks even if there was no course credit for such activities. 
With the assumption that the students will write on the exam booklet and write down relevant concepts only if they think it is helpful for
problem solving, multiple-choice exam can be a novel tool for assessment. It allowed us to observe students' problem solving strategies in a 
more ``native" form closer to what they really think is helpful for problem solving 
instead of what the professor wants them to write down or filling the page when a free-response
question is assigned with irrelevant equations and concepts with the hope of getting partial credit for the effort.

We decided to divide the students' work in the notebooks and exam-books into two categories: diagrams and scratchworks. 
The scratchworks included everything written apart from the diagrams such as equations, sentences, and texts.
Both authors of this paper agreed on how to differentiate between diagrams and scratchworks.
Instead of using subjectivity in deciding how ``good" the diagrams or scratchworks for each student for each of the 40 questions were,
we only counted the number of problems with diagrams drawn and scratchworks done by each student. For example, if a student drew diagrams
for 7 questions out of 40 questions and did scratchworks for 10 questions out of 40 questions, we counted it as 7 diagrams and 10 scratchworks. 

\vspace{-0.2in}
\section{Results}
\vspace{-0.1in}

Although no pretest was given to students, there is some evidence that, over the years, the evening section
of the course is somewhat weaker and does not perform as well overall as the daytime section of the course historically. 
The difference between the daytime and evening sections of the 
course could partly be due to the fact that some students in the evening section work full-time and take classes simultaneously.
For example, the same professor had also taught both sections of the course one year before the peer reflection activities were
introduced in evening recitations and thus all recitations for both sections of the course were taught traditionally that year. 
Thus, we first compare the averages of the daytime and evening sections before and after the peer reflection activities were instituted in
the evening recitation classes.  Table 1 compares the difference in the averages between the 
daytime and evening classes the year prior to the introduction of peer reflection (Fall 2006) 
and the year in which peer reflection was implemented in the evening recitation classes (Fall 2007). 
In Table 1, the p-values given are the results of t-tests performed between the daytime and evening classes.
Statistically significant difference (at the level of $p=0.05$) between groups only exists between the average midterm scores for the year 
in which peer reflection was implemented. 
The evening section scored lower on average than the daytime section on the final exam but the difference is not statistically significant
(p=0.112 for 2006 and p=0.875 for 2007), as indicated in Table 1. 
We note that while the midterm questions differed from year to year (since the midterms were returned to students and there was a possibility
that the students would share them), the final exam, which was not returned to students,
was almost the same both years.

\begin{table}[h]
\centering
\begin{tabular}[t]{|c|c|c|c|}
\hline
Daytime vs. & Daytime& Evening& p-value  \\[0.5 ex]
evening classes& means \%& means\% & \\[0.5 ex]
\hline
2006: midterm & 72.0&65.8 (non-PR)&0.101\\[0.5 ex]
\hline
2006: final exam & 55.7&52.7 (non-PR)&0.112\\[0.5 ex]
\hline
2007: midterm & 78.8&74.3 (PR)&0.004\\[0.5 ex]
\hline
2007: final exam & 58.1&57.7 (PR)&0.875\\[0.5 ex]
\hline
\end{tabular}
\vspace{0.1in}
\caption{Means and p-values for comparisons of the daytime and evening classes
during the year before peer reflection was introduced (Fall 2006) and during the year in which 
it was introduced (Fall 2007). The following were the number of students in each group: Fall 2006 daytime N=124, evening N=100, Fall 2007 daytime N=107,
evening N=93.}
\label{junk2}
\end{table}

\begin{table}[h]
\centering
\begin{tabular}[t]{|c|c|c|c|c|}
\hline
& Quest. & Traditional & PR group& p-value  \\[0.5 ex]
& type & group per & per & between  \\[0.5 ex]
& & student & student& groups  \\[0.5 ex]
\hline
$\#$ with& All & 7.0& 8.6& 0.003  \\[0.5 ex]
diagram& Quant. & 4.3& 5.1& 0.006  \\[0.5 ex]
& Concept& 2.7 & 3.5& 0.016  \\[0.5 ex]
\hline
$\#$ with& All & 20.2& 19.6& 0.496  \\[0.5 ex]
scratch& Quant. & 16.0 & 15.6 &0.401    \\[0.5 ex]
work& Concept& 4.2 & 4.0 &0.751  \\[0.5 ex]
\hline
\end{tabular}
\vspace{0.1in}
\caption{Comparison of the average number of problems per student with diagrams and scratchworks by the traditional group (N=107) and the PR group 
(N=93) in the final exam. The PR group has significantly more problems with diagrams than the traditional group. The average number of problems
with scratchworks per student in the two groups is not significantly different. 
There are more quantitative problems with diagrams drawn and scratchworks written than conceptual problems (at the level of $p=0.000$).}
\label{junk2}
\end{table}

The final exam which was comprehensive had 40 multiple-choice questions, half of which were quantitative and half were conceptual.
There was no partial credit given for drawing the diagrams or doing the scratchworks. One issue we investigated is whether the students considered the 
diagrams or the scratchworks to be beneficial and used them while solving problems, even though students knew that no partial credit was given for showing work.
As noted earlier, our assessment method involved counting the number of problems with diagrams and scratchworks. We counted any 
comprehensible work done on the exam notebook other than a diagram as a scratchwork. In this sense, quantifying the amount of scratchwork does not 
distinguish between a short and a long scratchwork for a given question. If a student wrote anything
other than a diagram, e.g., equations, the known variables and target quantities, an attempt to solve for unknown etc., 
it was considered scratchwork for that problem. Similarly, there was a diversity in the quality of diagrams the students
drew for the same problem. Some students drew elaborate diagrams which were well labeled while others drew rough sketches. Regardless of
the quality of the diagrams, any problem in which a diagram was drawn was counted.
We find that the PR group on average drew more diagrams than the traditional group.
Table 2 compares the average number of problems with diagrams or scratchworks by the traditional group and the PR group on the final exam. 
Tables 2 shows that, regardless of whether they belonged to the traditional group or the PR group, students were more likely to draw diagrams for 
the quantitative questions than for the conceptual questions. 

We also investigated whether the final exam score is correlated with the number of problems with diagrams or scratchworks
for the traditional group and the PR group separately (R is the correlation coefficient). The null hypothesis in each case is 
that there is no correlation between
the final exam score and the variable considered, e.g., the total number of problems with diagrams drawn.
We find that, for both the traditional group and the PR group, the students who had more problems with diagrams and scratchworks were
statistically (significantly) more likely to perform well on the final exam. 

\vspace{-0.15in}
\section{Discussion}
\vspace{-0.07in}

According to Chi~\cite{glaser}, students are likely to improve their approaches to problem solving and learn meaningfully from an intervention if both
of the following happen: I) students compare two artifacts, e.g., the expert solution and their own solution and realize their omissions,
 and II) they receive guidance to understand why the expert solution is better and how they can improve upon their
own approaches. The PR approach uses such a two tier approach in which students first identify that other student's solution may be
better than their own and then are guided by the UTAs/TA to reflect upon the various aspects of the ``winning" solutions. 

\vspace{-0.15in}
\begin{theacknowledgments}
\vspace{-0.07in}
We are grateful to NSF for financial support.
\end{theacknowledgments}

\bibliographystyle{aipproc}

\vspace{-0.15in}
\begin{thebibliography}{a}
\vspace{-0.1in}

\bibitem{edit1} E. Yerushalmi, C. Singh, and B. Eylon, AIP Conference Proceedings, Mellville NY {\bf 951}, 27-30 (2007).

\bibitem{singhgroup} C. Singh, Am. J. Phys. {\bf{73}}(5), 446-451 (2005).

\bibitem{mason} A. Mason $\&$ C. Singh, Am. J. Phys. {\bf 78(7)}, 748-754 (2010).

\bibitem{cog1} Collins, Brown $\&$ Newman, in Resnick (Ed.),
Essays in honor of R. Glaser, Lawrence Erlbaum., 453-494, (1989). 

\bibitem{glaser} M. T. H. Chi, in R. Glaser (Ed.). Advances in Instructional Psychology, Lawrence Erlbaum, 161-238 (2000). 

\end{thebibliography}

\end{document}